\documentclass[12pt,preprint]{aastex}

\slugcomment{Submitted to Astrophysical Journal}

\shorttitle{3D Delayed-Detonation Model of SN~Ia}
\shortauthors{Gamezo, Khokhlov, \& Oran}

\begin{document}

\title{Three-Dimensional Delayed-Detonation Model \\ of Type Ia Supernova}

\author {Vadim N. Gamezo\altaffilmark{1}, 
        Alexei M. Khokhlov\altaffilmark{2}, 
    and Elaine S. Oran\altaffilmark{1}}

\altaffiltext{1}{
Laboratory for Computational Physics and Fluid Dynamics, 
Naval Research Laboratory, Washington, D.C. 20375, 
gamezo @lcp.nrl.navy.mil, oran @lcp.nrl.navy.mil
}
\altaffiltext{2}{
Department of Astronomy and Astrophysics, University of Chicago,
Chicago, IL 60637, \hbox{ajk @oddjob.uchicago.edu}
}

\begin{abstract}

We study a Type~Ia supernova explosion using large-scale
three-dimensional numerical simulations based on reactive fluid
dynamics with a simplified mechanism for nuclear reactions and energy
release.  The initial deflagration stage of the explosion involves a
subsonic turbulent thermonuclear flame propagating in the gravitational
field of an expanding white dwarf. The deflagration produces an
inhomogeneous mixture of unburned carbon and oxygen with
intermediate-mass and iron-group elements in central parts of the
star.  During the subsequent detonation stage, a supersonic detonation
wave propagates through the material unburned by the deflagration.  The
total energy released in this delayed-detonation process,
$(1.3-1.6)\times10^{51}$~ergs, is consistent with a typical range of
kinetic energies obtained from observations.  In contrast to the
deflagration model that releases only about $0.6\times10^{51}$~ergs,
the delayed-detonation model does {\it not} leave carbon, oxygen, and
intermediate-mass elements in central parts of a white dwarf. This
removes the key disagreement between three-dimensional simulations and
observations, and makes a delayed detonation the mostly likely mechanism
for Type~Ia supernova explosions.

\end{abstract}

\keywords{supernovae: general ---  hydrodynamics ---  nuclear reactions, 
nucleosynthesis, abundances }

\section{Introduction}

One-dimensional (1D) numerical models of Type Ia supernovae (SN~Ia)
have been extensively used to test general ideas about possible
explosion mechanisms \cite{ref2,ref3,ref4,ref5,ref6,ref7}.  In
particular, 1D models have ruled out the possibility of a thermonuclear
detonation, a supersonic shock-induced combustion mode, as a sole
mechanism for SN~Ia explosions.  A detonation propagating through a
high-density carbon-oxygen white dwarf (WD) produces mostly Ni and
almost none of the intermediate-mass elements, such as Ne, Mg, Si, S,
and Ca, that are observed in SN~Ia spectra.  One-dimensional models
have also shown that a detonation can produce intermediate-mass
elements if it propagates through a low-density WD preexpanded during
the initial deflagration stage of the explosion.  These
delayed-detonation models
\cite{ref8,ref9,ref10,ref11,ref12,ref13,ref14}, which have a
deflagration-to-detonation transition (DDT) at some stage of the
thermonuclear explosion, are the most successful in reproducing
observed characteristics of SNe~Ia. Many crucial physical details,
however, including the mechanism of DDT and the turbulent flame
structure, are missing by definition from 1D models because SN~Ia
explosions are intrinsically three-dimensional (3D) phenomena.

Recent 3D numerical simulations of thermonuclear deflagrations in a WD
\cite{ref15,ref16,ref17,ref18} demonstrated the importance of 3D
effects.  In particular, they have shown that the development of a
turbulent thermonuclear flame in the gravitational field of a WD allows
funnels of unburned and partially burned material to remain in the
vicinity of the WD center until the end of the explosion. This would
produce distinct signatures of low-velocity carbon, oxygen, and
intermediate-mass elements in SN~Ia spectra.  As the observed spectra
do not show these signatures, the predictions of the pure deflagration
model are inconsistent with observations. This inconsistency can be
resolved if the turbulent flame triggers a detonation that can burn the
remaining material near the WD center \cite{ref15,ref18,ref1}.  
Here, we report results numerical simulations of SN~Ia explosions based
on a 3D delayed-detonation model.  This work extends and clarifies the
brief letter \cite{ref1} by providing a detailed description of the
model and the results, including numerical convergence studies and
extended data analysis.

\section{Input Physics and Numerical Implementation}

The numerical model \cite{ref15,ref18} is based on reactive Euler
equations that include gravity terms and are coupled with an equation
of state for a degenerate matter and a simplified kinetics of energy
release. The equations are integrated on a Cartesian adaptive mesh
using an explicit, second-order, Godunov-type numerical scheme. The
model describes compressible fluid dynamics on large scales in an
exploding WD including the expansion of a star, Rayleigh-Taylor (RT)
and Kelvin-Helmholtz (KH) instabilities, turbulence, pressure waves,
shocks, and detonations.  The nuclear kinetics is approximated by a
four-equation mechanism \cite{ref8,ref15} that describes the energy
release, consumption of carbon, NSE and NSQE relaxations, and
neutronization.  Appendix~A describes details of the nuclear kinetics
approximation and explains how it can be used to estimate cumulative
mass fractions of major groups of elements.

The model is able to reproduce the two different regimes of the
thermonuclear burning in a WD, a subsonic deflagration and a supersonic
detonation.  These regimes differ by the mechanism of propagation of
the reaction wave: a deflagration involves heat conduction or turbulent
mixing, and a detonation involves shock compression. For both regimes,
the energy is released by the same network of thermonuclear reactions,
and the physical thickness of the reaction front strongly depends on
density.  It can be up to 12 orders of magnitude less than the WD
radius $R_{WD}$ for deflagrations \cite{ref19,ref20} and up to 10
orders of magnitude less than $R_{WD}$ for detonations
\cite{ref21,ref22}.  Since the large-scale simulations described here
do not resolve lengthscales smaller than $10^{-3}R_{WD}$, the reaction
fronts at high densities are still unresolved. We explicitly resolve
only parts of the reaction zone associated with NSE relaxation that
become very large at low densities and cause an incomplete burning that
produces Si and other intermediate-mass elements.

Unresolved reaction fronts are treated differently for deflagrations
and detonations. For deflagrations, the flame is advanced using a
flame-capturing algorithm described in Appendix~B that ensures the
flame propagation with a prescribed speed.  The flame speed is provided
by an additional subgrid model \cite{ref15,ref18} that takes into
account physical processes at scales smaller than the computational
cell size.  In particular, it assumes that turbulent burning on small
unresolved scales is driven by the gravity-induced RT instability.
This approach works only if we explicitly resolve enough scales to
ensure that the resolved turbulent flame structure corresponds to the
flame properties build into the subgrid model.  Extensive numerical
resolution tests have shown that the minimum computational cell size
$\Delta x_{min} = 2.6\times10^5$~cm used here for the deflagration
stage is adequate for this type of simulation \cite{ref18}.

For detonations, a front-capturing algorithm and a subgrid model are
not needed because the quasi-steady-state velocity of a detonation wave
and the equilibrium composition of detonation products do not depend on
the kinetics of energy release. We thus can artificially limit the
maximum reaction rate in order to increase the reaction-zone thickness
to several computational cells and ensure the numerical stability.  We
implement this by requiring that the mass fraction of carbon should not
change by more than 0.1 per timestep. The resulting artificially thick
detonation wave described only by reactive Euler equations maintains
the correct velocity and burns the material to the equilibrium products
regardless of the computational cell size.  Resolution tests performed
for the detonation stage for $\Delta x_{min} = 10.4\times10^5$~cm and
$5.2\times10^5$~cm show no substantial differences in results (see also
Fig.~5 and related discussion below).  Transient detonation phenomena,
such as detonation initiation and extinction, cannot be correctly
reproduced without resolving the structure of the detonation wave, and
are not included in the simulations. 

\section{Deflagration Stage}

The initial conditions for the deflagration stage model a $1.4M_\odot$
WD in hydrostatic equilibrium with initial radius
$R_{WD}=2\times10^8$~cm, initial central density $\rho_c=
2\times10^9$~g/cm$^3$, spatially uniform initial temperature
$T=10^5$~K, and uniform initial composition with equal mass fractions
of $^{12}$C and $^{16}$O nuclei. The burning was initiated at the
center of WD by filling a small spherical region at $r<0.015R_{WD}$
with hot reaction products without disturbing the hydrostatic
equilibrium.  We model one octant of the WD assuming mirror symmetry
along the $x=0$, $y=0$ and $z=0$ planes. The computational domain is a
cube with a side of $x_{max}=5.35\times10^8$~cm.

The development of the thermonuclear flame was described in detail
earlier \cite{ref18} and is shown in Fig.~1. The initially spherical
flame propagates outwards with a laminar velocity $\sim$$100$~km/s,
becomes distorted due to the RT instability, and forms multiple plumes
at different scales.  The flame distortions at small unresolved scales
are taken into account by the turbulent subgrid model that controls the
normal flame speed. As the flame grows and the star expands, the normal
flame speed remains close to $\sim$$100$~km/s, which is small compared
to the sound speed $4,000-10,000$~km/s ahead of the flame.  On resolved
scales, the flame forms a dynamic convoluted surface penetrated in all
directions by very irregular  funnels of unburned material.  Buoyancy
causes the hot, burned, low-density material inside the flame plumes to
rise towards the WD surface. The same gravitational forces also pull
the cold, unburned, high-density material between the plumes down
towards the center.  This material continues to burn, but it will not
burn out completely as long as convective flows supply fresh unburned
material from outer layers.

At high densities, carbon and oxygen burning produces mostly Ni and
other iron-group elements. The energy released increases the
temperature of burned material to $\sim$$10^{10}$~K and causes the WD
to expand. When the density of unburned material becomes lower than
$(1-5)\times10^7$~g/cm$^3$, the deflagration begins to produce Si and
other intermediate-mass elements.  These elements form everywhere along
the flame surface, starting from the outer layers reached by the large
plumes, and continuing around unburned funnels in the inner parts of
the WD.  The 3D distribution of elements predicted at 1.94~s after
ignition is shown in Fig~2d.

The expansion eventually quenches the burning when the density of
unburned material drops below $\simeq$$10^6$~g/cm$^3$.  The unburned
carbon and oxygen remaining between the flame plumes and
intermedi\-ate-mass elements that form at low densities at different
radii are likely to survive the explosion. As we have shown
\cite{ref15,ref18}, this makes predictions of the 3D deflagration model
inconsistent with observed spectra of SN~Ia.

\section{Detonation Stage}

The disagreement between predictions from the pure deflagration
simulation and observations strongly suggests that the turbulent flame
in SN~Ia triggers a detonation.  The process of DDT involves events
occuring at small scales that are comparable to the detonation wave
thickness, and, thus, cannot be directly modeled in large-scale
simulations.  To study the effects of a detonation, we therefore assume
a time and a location for DDT. (A similar approach has been used
previously in 1D \cite{ref8,ref9,ref10,ref11,ref12,ref13} and 2D
\cite{ref23,ref24,ref25} delayed-detonation models.)
We use the deflagration results as initial conditions, and impose a
hot spot to ignite the detonation.  The time and location for the
detonation initiation are parameters that can be varied and optimized.
Here, we explore the three cases defined in Table~\ref{table1}.

\begin{table} [h]
\caption{\label{table1}
Parameters of a WD at the time detonation initiation ($t^{di}$) for
delayed-detonation cases computed:  burned mass fraction ($f_b^{di}$),
density of unburned material near the WD center ($\rho^{di}$), WD
radius ($R_{WD}^{di}$), maximum distance from the flame surface to the
WD center ($R_f^{di}$), distance from the initiation point to the WD
center ($R^{di}$).  Symmetry boundary conditions imply that the
off-center initiation in case (b) occurs at two points.}
\medskip
\begin{tabular}{ccccccc}
\tableline\tableline
Case & $t^{di}$ & $f_b^{di}$ & $\rho^{di}$     & $R_{WD}^{di}$
                             & $R_f^{di}$      & $R^{di}$ \\
     &          &            &$10^8$           & $10^8$ 
                             &$10^8$           & $10^8$   \\
     &       s  &            &       g/cm$^3$  &        cm 
                             &       cm        &        cm\\
\tableline
 a   & 1.62     & 0.33       & 2.5  & 3.1  & 2.3 & 0 \\
 b   & 1.62     & 0.33       & 2.5  & 3.1  & 2.3 & 1 \\
 c   & 1.51     & 0.25       & 4.4  & 2.6  & 1.7 & 0 \\
\tableline
\end{tabular}
\end{table}

The frames in the left column of Fig.~1 show the detonation propagation for
case (a), which corresponds to central detonation initiation at 1.62~s
after the beginning of the deflagration. By that time, the density of
unburnt material near the center has dropped to
$2.5\times10^8$~g/cm$^3$.  A detonation at this density produces mostly
Ni and propagates outwards at $\sim$$12,000$~km/s, which is comparable
to the expansion velocities induced by subsonic burning. When the
detonation reaches unburned material with density below
$(1-5)\times10^7$~g/cm$^3$, it begins to produce intermediate-mass
elements. Different parts of the detonation front that exit different
funnels collide with each other, coalesce, and eventually reach the
surface of the star.

The detonation transforms all carbon and oxygen in central parts the WD
into iron-group elements, and produces intermediate-mass elements in
outer layers. This drastically changes the distribution of nuclei
compared to that produced by  the pure deflagration.  Figure~2a shows
that funnels of unburned carbon and oxygen disappear from central parts
of the WD. Iron-group elements form a distinct core surrounded by a
layer of intermediate-mass nuclei. Angle-averaged mass fractions of the
main elements calculated for the deflagration and the
delayed-detonation models as functions of distance from the WD center
are compared in Fig.~3. 

Similar results were obtained for the delayed-detonation case (c) with
earlier central initiation at 1.51~s. The detonation starts in the
material with density $4.4\times10^8$~g/cm$^3$ propagates outwards, and
almost completely covers the flame plumes by 1.82~s, as shown in the
righ column of Fig.~1. By that time, that corresponds to Figs.~2c, 3c,
and 4c, the shock already reached the WD surface, burned almost all of
the carbon, and left unburned oxygen in low-density outer layers.  The
integral mass distribution of the main elements in Fig.~4 shows that 
case (c) produced more iron-group elements than case (a) because the
detonation propagated through higher-density material. Figure~2c shows
that iron-group elements form a more compact core because the flame
plumes had less time to develop. Intermediate-mass elements also form a
more distinct shell with a higher Si peak in Fig.~3c. Thus, the earlier
detonation initiation results in more symmetrical, shell-like
distribution of elements.

Cases (a) and (c) were computed with two different numerical
resolutions with $\Delta x_{min} = 10.4\times10^5$~cm and
$5.2\times10^5$~cm. Total energies for these cases are shown in Fig.~5
as functions of time. Dotted lines correspond to the lower resolution
and show a slightly lower energy release compared to high-resolution
results. The difference (6\% for the case (a) and 3\% for the case (c))
is insignificant and related to small pockets of unburned material that
remain behind the shock propagating through the turbulent flame in
low-resolution simulations.

The delayed-detonation case (b) was computed only with the low
resolution ($\Delta x_{min} = 10.4\times10^5$~cm). This case is
interesting because it shows asymmetrical effects created by an
off-center detonation initiation.  The simulation results for this case
show a moderate asymmetry in final composition.  The asymmetry is
limited because we calculate only one octant of a WD and impose mirror
boundary conditions. The degree of asymmetry would increase if the
simulations were performed for a full star. Then the second
mirror-reflected spot for detonation initiation would be eliminated.
The simulation results indicate that, during the period of detonation
propagation, the density of the expanding unreacted material ahead of
the shock can decrease by an order of magnitude compared to its value
at the end of the deflagration stage.  Because the detonation burns
material to different products at different densities, it can create a
large-scale asymmetry in composition if it starts far from the WD
center. A similar conclusion based on 2D simulations was made by
\cite{ref25}.  Three-dimensional simulations of the deflagration stage
\cite{ref15,ref26} also show that a developing flame, unrestricted by
mirror boundaries, can move away from the WD center, thus creating a
large-scale asymmetry at very early stages of the explosion.

\section{Discussion and Conclusions}

The first result that can be compared to observations is the energy
released by the explosion.  Total energies for all three
delayed-detonation cases and the deflagration model are compared in
Fig.~5. The total energy $E_{tot}$ here is the difference between the
energy released by thermonuclear reactions and the binding energy of
the star. Eventually $E_{tot}$ will be transformed into kinetic energy
of expanding material that can be measured in observations of SN~Ia.
Figure~5 shows that the total energy predicted by delayed-detonation
models, $(1.3-1.6)\times10^{51}$~ergs, is much higher than the total
energy predicted by the deflagration model
$\sim$$0.6\times10^{51}$~ergs. The reason for this is that the
deflagration is able to burn only about a half of the WD mass. The rest
of the material expands to the densities below $\simeq$$10^6$~g/cm$^3$
that do not support the thermonuclear burning.  A detonation propagates
faster and burns almost all of the WD mass before the material expands
to low densities.  A small difference in energy curves plotted in
Fig.~5 for the delayed-detonation cases (a) and (b) indicates that the
off-center detonation initiation in this model has only a minor effect
on the energy release.  Case (c), with earlier detonation initiation,
resulted in a faster explosion that released 15\% more energy.  The
total energy released by the delayed-detonation models is in agreement
with a typical range $(1-1.5)\times10^{51}$~ergs obtained from SN~Ia
observations \cite{ref7}.

Another result that can be compared to observations is the total mass
of radioactive $^{56}$Ni that provides the energy source for the
observed luminosity of SN~Ia.  The simplified kinetic model we use in
these simulations does not provide the exact concentration of $^{56}$Ni,
but it gives a cumulative concentration of iron-group elements, most of
which is $^{56}$Ni.  By the end of each simulation described here, all
iron-group elements that could form during the explosion have already
formed. Even though some material is still burning and releasing energy
for cases (a), (b), and (d) (see Fig.~5), the burning occurs at low
densities and produces mostly intermediate-mass elements.  The total
mass of iron-group elements created by the explosion (see Fig.~4) is
0.78, 0.73, and 0.98 solar masses ($M_\odot$) for delayed-detonation
cases (a), (b), and (c), respectively. The mass of $^{56}$Ni estimated
from observational data is about $0.6M_\odot$ for a typical SN~Ia
\cite{ref27}, and is in agreement with the total mass of iron-group
elements produced by delayed-detonation models. For the deflagration
model, the total mass of iron-group elements is only 0.47 $M_\odot$,
which is insufficient to account for the luminosity of a typical
SN~Ia.

Distributions of carbon, oxygen, and intermediate-mass elements
predicted by the models also can be checked against observations.  In
the simulations, detonations burn all carbon and oxygen in inner parts
of WD. Oxygen remains unburned in outer layers, which expand to
densities below $\simeq$$10^6$~g/cm$^3$ before the detonation reaches
them. Carbon is likely to remain unburned for densities below
$(1-3)\times10^5$~g/cm$^3$.  These unburned carbon and oxygen in outer
layers would produce spectral signatures only in the high-velocity
range.

There are ways in which a delayed detonation can leave small amounts of
carbon and oxygen in inner parts of WD. For example, a detonation
propagating through a thin, sinuous funnel of unburned material can
fail if the funnel makes a sharp turn. This detonation failure occurs
as a result of detonation diffraction process that is well-known for
terrestrial systems.  A developing turbulent flame can also disconnect
some funnels from the rest of the unburned material, thus creating
unburned pockets that cannot be directly reached by a detonation wave.
Some of these pockets may not ignite when strong shocks generated by
detonations reach them.  The cellular structure of thermonuclear
detonations in carbon-oxygen matter \cite{ref22}, and the ability of
cellular detonations to form pockets of unburned material that extend
far behind the 1D reaction zone, can also contribute into incomplete
burning. All these phenomena occur at lengthscales comparable to the
reaction zone thickness that are not resolved in large-scale
simulations reported here, and thus require additional studies.

There have recently been efforts to detect low-velocity carbon in SN~Ia
spectra that could result from the funnels of unburned material near
the WD center \cite{ref28}. The results \cite{ref28} indicate that
carbon can be present at velocities 11,000~km/s.  Even though this
velocity is much lower than 20,000-30,000~km/s usually attributed to
carbon in SN~Ia spectra \cite{ref29,ref30,ref31}, it is still too high
for the ejecta formed from central parts of a WD. For carbon and
oxygen, spectral signatures are difficult to observe, and estimated
velocities of these elements are subject to large uncertainties.
Intermediate-mass elements, however,  produce distinct spectral lines
and their velocities are well defined. The minimum observed velocities
for intermediate-mass elements \cite{ref28,ref32} are large enough
($\sim$$10,000$~km/s for Si) to rule out the presence of these elements
near the WD center, as is predicted by the deflagration model.  A
discussion on this subject can also be found in the recent article by
\cite{ref33}.

Figures 2-4 show that, in contrast to the 3D deflagration model,
the 3D delayed-detonation model of SN~Ia explosion does {\it not} leave
carbon, oxygen, and intermediate-mass elements in central parts of a
WD. This removes the key disagreement between simulations and
observations, and makes the 3D delayed detonation a promising mechanism
for SN~Ia explosion.  Further analysis of 3D delayed detonations on
large scale requires 3D radiation transport simulations to produce
spectra, and a detailed comparison between the calculated and observed
spectra of SN~Ia for different initiation times and locations. The
uncertainty in detonation initiation can only be eliminated by solving
the DDT problem that requires resolving physical processes at small scales.

\acknowledgments

This work was supported in part by the NASA ATP program
(NRA-02-OSS-01-ATP) and by the Naval Research Laboratory (NRL) through
the Office of Naval Research. Computing facilities were provided by
the DOD HPCMP program. We would like to thank Peter H\"oflich
and J. Craig Wheeler for useful discussions.

\appendix

\section{Nuclear Kinetics}       

Nuclear reactions involved in the thermonuclear burning of the
carbon-oxygen mixture \cite{refS1,refS2,refS3} can be separated into three
consecutive stages responsible for the energy release. First,  the
$^{12}$C~+~$^{12}$C reaction leads to the consumption  of $C$ and
formation of mostly Ne, Mg, protons, and $\alpha$-particles.  Then
begins the nuclear statistical quasi-equilibrium (NSQE) relaxation,
during which $O$ burns out and Si-group (intermediate mass) elements
are formed.  Finally, Si-group elements are converted into the Fe-group
elements and the nuclear statistical equilibrium (NSE) sets in. The
reaction time scales $\tau_{C} << \tau_{nsqe} <<  \tau_{nse}$
associated with these stages strongly depend on temperature and density
and may differ from one another by several orders of magnitude
\cite{refS4,refS5,refS6,ref21,ref22}.

The full nuclear reaction network includes hundreds of species that
participate in thousands of reactions.  Integration of this full
network is too time-consuming to be used in multidimensional numerical
models. Therefore, we used a simplified four-equation kinetic scheme
\cite{ref8,ref15} that describes the energy release, NSE and NSQE
relaxations, and neutronization of NSE matter.

The kinetic equation for the mole  fraction of carbon $Y_C$
\begin{equation} \label{eqS1}
{dY_C\over dt} = -\rho A(T_9) \exp(-Q/T_{9a}^{1/3}) Y_C^2~, 
\end{equation}
describes carbon consumption  through the major reactions $^{12}C \,
(\,{^{12}C},\,p) \, ^{23}Na \, (\,p,\gamma) \, ^{24}Mg$ and 
$^{12}C\,(\,{^{12}C},\,{^4He})\,^{20}Ne$  with the branching ratio
$\simeq$1. Here $Q=84.165$, $T_{9a} = T_9/(1+0.067T_9)$, where
$A(T_9)$ is a known function \cite{refS1}.

Most of the nuclear energy is released during the carbon exhaustion
stage and  the subsequent synthesis of the Si-group nuclei. The energy
release or consumption due to the transition from Si-group to Fe-group
nuclei (NSE relaxation) is  less than 10\%. Therefore, the nuclear
energy release rate is approximated as
\begin{equation} \label{eqS2}
{dq_n \over dt} = -Q_C {dY_C \over dt} +  { q_{nse} - q_n \over \tau_{nsqe} }~, 
\end{equation}
where $q_n(t)$ is the binding energy of nuclei per unit mass, $Q_C =
4.48\times 10^{18}$ ergs g$^{-1}$ mol$^{-1}$ is the energy release due
to carbon burning described by Eq.~(\ref{eqS1}), and $q_{nse}(\rho,T,Y_e)$ is the
binding energy of matter in the state of NSE.

The  equation 
\begin{equation} \label{eqS3}
{d\delta_{nse}\over dt} = { 1 - \delta_{nse} \over \tau_{nse}}~,    
\end{equation}
traces the onset of NSE, where $\delta_{nse} = 0$ in the unburned
matter and $\delta_{nse} = 1$ in the NSE products. Intermediate mass
elements  are expected where $Y_C(t=\infty) \simeq 0$ and
$\delta_{nse}(t=\infty) < 1$.  Fe-group elements are expected  where
$Y_C(t=\infty) \simeq 0$ and $\delta_{nse}(t=\infty) \simeq 1$.
The `e-folding' NSQE and NSE timescales  
\begin{equation} \label{eqS4}
\tau_{nsqe} = \exp \left( 149.7/T_9 -39.15 \right)~{\rm s}~, ~~~
\tau_{nse}  = \exp \left( 179.7/T_9 -40.5  \right)~{\rm s}~, 
\end{equation}
approximate the results of the detailed calculations of carbon burning. 

Neutronization is described by the equation for the electron mole
fraction $Y_e$,
\begin{equation} \label{eqS5}
{dY_e \over dt} = - R_w(\rho,T,Y_e)~.                            
\end{equation}

The corresponding term describing  neutrino energy losses, $-\dot
q_w(\rho,T,Y_e)$, is added to the nuclear energy release rate used in
reactive Euler equations.  Values of $q_{nse}$, $\dot q_w$, and $R_w$
were computed assuming  NSE distribution of individual nuclei, as
described in \cite{ref8}. Recently, there has been an important
development in theoretical computations  which shows significantly
smaller  electron capture and $\beta$-decay rates in stellar matter
(\cite{refS11} and references therein).  Following \cite{refS12},  this effect
was  approximately taken  into account by decreasing $\dot q_w$ and
$R_w$ by a factor of $5$. This should  be sufficient to  account for
changes in $q_{nse}$ and in the corresponding  nuclear energy release
$q_{nse} - q_n(t=0)$ caused by variations of $Y_e$.

The original  kinetic scheme \cite{ref8} also contained the equation for
the mean ion mole fraction $Y_i$. Here we neglect variations of $Y_i$
due to nuclear reactions since ions make a small contribution to the
equation of state.  A constant $Y_i=0.07$ was used instead, which is an
average of $Y_i$ in the unburned and typical NSE matter in the SN~Ia
explosion conditions.

Despite its simplicity, the kinetic scheme adequately describes all
major stages of the thermonuclear burning in a carbon-oxygen WD. In
particular, it takes into account  the important effect of energy
release or absorbtion caused by changes in NSE composition when matter
expands or contracts (these changes happen on a quasi-equilibrium
rather than on the equilibrium timescale).  When the WD expands, this
effect  adds $\simeq$50\% to the energy initially released by burning
at high densities. Using even the simplest 13-species $\alpha$-network
to account for this effect would be prohibitively expensive in 3D
simulations of SN~Ia explosions. The same kinetic scheme was used in
\cite{ref15,ref18}  for the deflagration stage, but the nuclear energy
release was overestimated by about 17\% due to an incorrect numerical
implementation. Here use a corrected model that produces slightly
slower and less energetic deflagrations.

The mass fraction of carbon $X_{C}$ and the NSE progress variable
$\delta_{nse}$ provided by the kinetic scheme were used to estimate
cumulative mass fractions of iron-group elements $X_{Ni}$,
intermediate-mass elements $X_{Si}$, elements from Mg to Ne $X_{Mg}$,
and a mass fraction of oxygen $X_{O}$, assuming that $X_{C} + X_{O} +
X_{Mg} + X_{Si} + X_{Ni} = 1$. The estimation scheme summarized in
Table~\ref{tableA1} is based on the analysis of the reaction zone structure of a
1D detonation wave calculated in \cite{ref21} with a detailed nuclear
kinetics.  The estimated mass fractions were not involved in fluid
dynamics simulations and used only for the analysis of the simulation
results.

\begin{table}[h]
\caption{ \label{tableA1}
Estimation scheme for mass fractions of main elements.  
}
\smallskip \noindent
\begin{tabular*}{\textwidth}{@{\extracolsep{\fill}}lcccc}
\tableline\tableline
\\
Mass          & & $X_C>\delta_s$    & $X_C<\delta_s$           &  $X_C<\delta_s$      \\
fraction      & &                   & $\delta_{nse}<\delta_s$  &  $\delta_{nse}>\delta_s$  \\
\\
\tableline
\\
$X_O   $      & & $ X_O^0         $ & $X_O^0 (1-\delta_{nse}/\delta_s)           $                  & 0                  \\ 
$X_{Mg}$      & & $(X_C^0 - X_C)/2$ & $(X_C^0 - X_C) (1-\delta_{nse}/\delta_s)/2 $                  & 0                  \\ 
$X_{Si}$      & & $ X_{Mg}        $ & $(X_C^0 - X_C)/2 + (1-(X_C^0 + X_C)/2) \delta_{nse}/\delta_s$ & $1-\delta_{nse}$        \\ 
$X_{Ni}$      & & 0                 & 0                                                             & $\delta_{nse}-X_C$      \\ 
\\
\tableline
\end{tabular*}
\tablenotetext{}{
$X_C^0$ and $X_O^0$ are initial mass fractions of carbon and oxygen, $\delta_s=0.001$. 
}
\end{table}


\section{Flame propagation}

The flame speed $S$ in our simulations is provided by a subgrid model
\cite{ref15,ref18} that takes into account physical processes at scales
smaller than the computational cell size. The flame is advanced using a
flame-capturing technique which mimics a flame propagating with a
prescribed normal speed \cite{refS14,ref15,ref18}.  A scalar variable
$f$, such that $f=0$ in the unburned matter and $f=1$ in the material
which has passed through the flame, obeys a reaction-diffusion equation
\begin{equation} \label{eqS6}
{{\partial f}\over{\partial t}} + {\bf U}\cdot\nabla f = K \nabla^2f + R~, 
\end{equation}
with artificial reaction and diffusion coefficients 
\begin{equation} \label{eqS7}
K  =  {\rm const},~~~~~
R  = \left\{ \begin{array}{ll}
      C={\rm const.,} & {\rm if}~~ f_0 \leq f \leq 1; \\
      0,              & {\rm otherwise,}
   \end{array}  \right.  \\
\end{equation}
where $f_0=0.3$. Equation (\ref{eqS6}) has a solution $f(x-St)$ which
describes a reaction front that propagates with the speed $S=
(KC/f_0)^{1/2}$ and has a thickness $\delta \simeq (K/C)^{1/2}$ (see
Appendix of \cite{refS14}).  If $K$ and $C$ in Eq.~(\ref{eqS7}) are set
to

\begin{equation} \label{eqS8}
K = S  (\beta \Delta x) \sqrt{f_0} ~,~~~~~
C = {S \over ( \beta \Delta x) } \sqrt{f_0}~,                        
\end{equation}
with $\beta={\rm const}$, the front propagates with a prescribed speed
$S$ and  spreads onto several  computational cells of size $\Delta x$.
The choice of  $\beta=1.5$ spreads the flame  on $\simeq$3-4 cells.
Making the front narrower  is not practical, since the fluid dynamics
algorithm spreads contact discontinuities on $\simeq$4 cells. 

To couple Eq.~(\ref{eqS6}) and the nuclear kinetic scheme
described in Appendix~A, the energy-release rate inside the front is 
defined as
\begin{equation} \label{eqS9}
\dot q =  \,q_f {df\over dt}~,                                       
\end{equation}
where the nuclear energy release inside the front $q_f$ is calculated
depending on the density as: 
\begin{equation} \label{eqS10}
q_f = \left\{ \begin{array}{ll}
       q_{nse} - q_n(0)    & {\rm if}~~ \rho > 2\times 10^7~g/cm^3  \\
       Q_CY_C(t=0)         & {\rm if}~~ \rho < 5\times 10^6~g/cm^3  \\
{\rm linear~interpolation} & {\rm if~~ otherwise}                   \\
      \end{array}  \right.  \\
\end{equation}
The carbon mole fraction $Y_C$ inside the front is decreased in
proportion to the increase of $f$.  Thus, at high densities, both
carbon consumption and the NSQE relaxation take place inside the flame
front. At low densities where the NSQE stage of burning is slow,  NSQE
relaxation takes place outside the flame front.  This flame-capturing
technique advances the reaction front relative to the fuel with the
speed practically independent of the orientation of the front on the
mesh, fluid motions, and resolution.

\vfill\eject


\epsscale{0.9}
\begin{figure}           
\caption{
Development of a turbulent thermonuclear flame (colored surface) and a
detonation (gray surface) in a carbon-oxygen WD.  Numbers show time in
seconds after ignition.  Central column shows the deflagration stage.
Left and right columns correspond to delayed-detonation cases (a)
(detonation starts at 1.62~s) and (c) (detonation starts at 1.51~s),
respectively.  Flames at 0.30, 0.61, 0.90, and 1.20~s are plotted at
the same scale. Further flame growth is shown by the color scale that
changes with distance from the flame surface to the WD center.
$x_{max}=5.35\times10^8$~cm.
\label{fig1}}
\end{figure}

\epsscale{1.0}
\begin{figure}           
\caption{
Concentration field in the exploding WD computed for deflagration (d)
and delayed-detonation (a,c) models defined in Table~\ref{table1}.
Times are 1.94, 1.94, and 1.82~s for cases d, a, and c, respectively.
The color map shows the average atomic number $A$ of the material for
$x=0.05$, $y=0.05$, and $z=0.05$ planes.  The coordinate grid is spaced
by $0.2x_{max}$.  $A=\sum X_i A_i$, where $X_i$ are mass fractions of
C, O, Mg, Si, Ni defined in Appendix~A.
\label{fig2}}
\end{figure}

\epsscale{0.5}
\begin{figure}           
\plotone{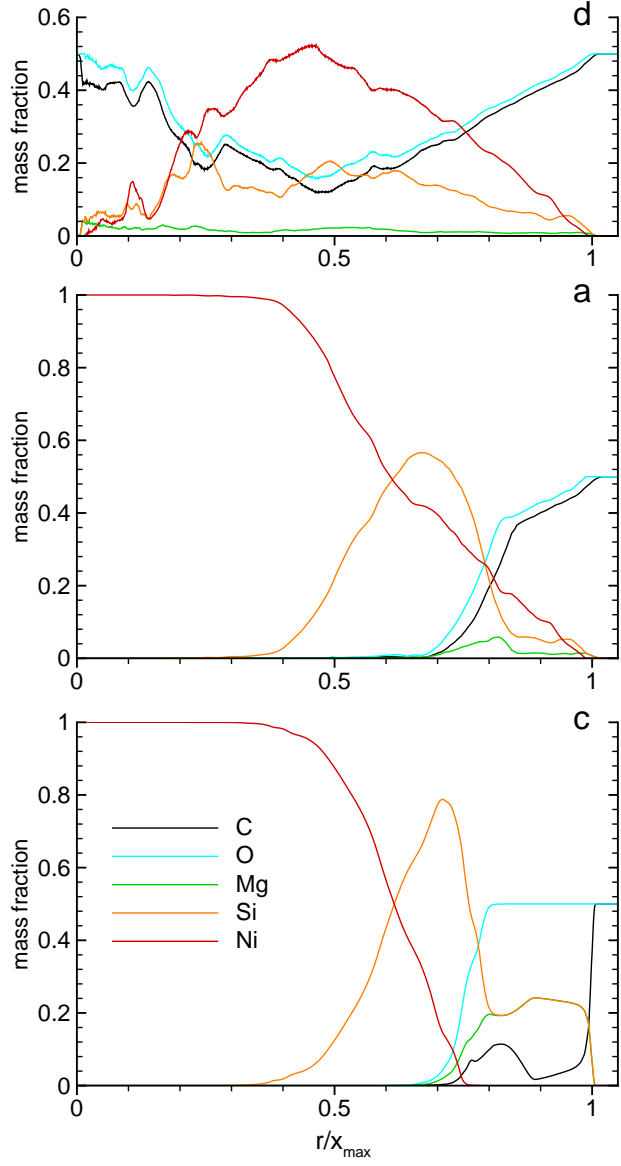}
\caption{
Angle-averaged mass fractions of the main elements as functions of
scaled distance from the WD center produced by the deflagration (d) and
delayed-detonation (a,c) models defined in Table~\ref{table1}.  Times
are 1.94, 1.94, and 1.82~s after the beginning of the explosion for
cases d, a, and c, respectively.  $x_{max}=5.35\times10^8$~cm.
\label{fig3}}
\end{figure}

\begin{figure}           
\plotone{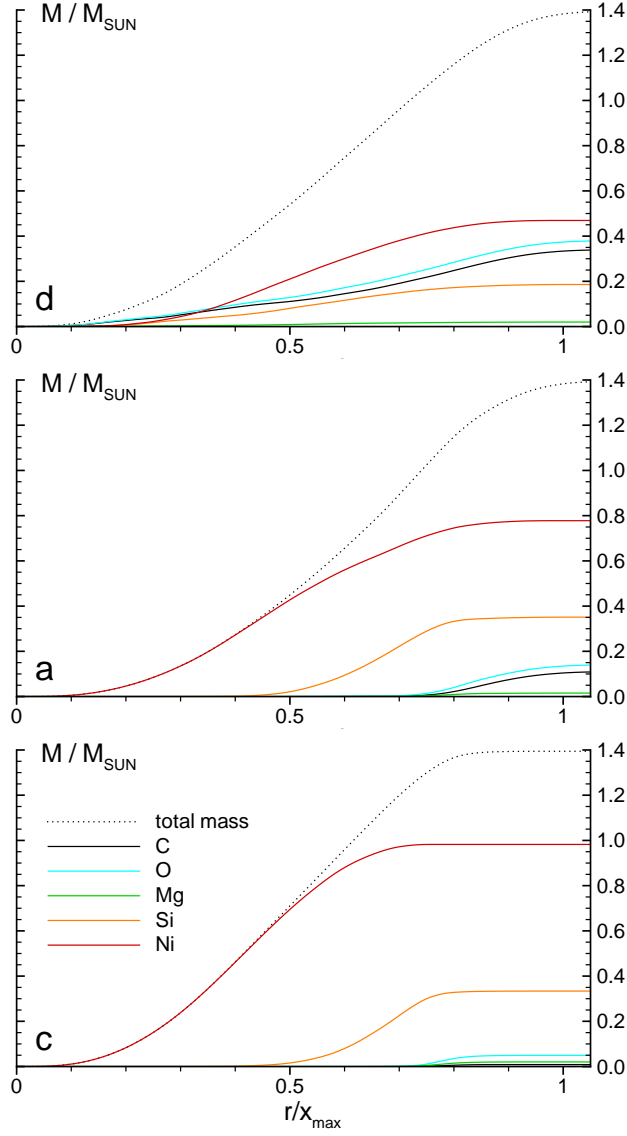}
\caption{
Integral mass distribution of the main elements in the exploding WD
produced by the deflagration (d) and delayed-detonation (a,c) models
defined in Table~\ref{table1}. Times are 1.94, 1.94, and 1.82~s after
the beginning of the explosion for cases d, a, and c, respectively.
The mass is scaled by the solar mass, the radius is scaled by the
computational domain size $x_{max}=5.35\times10^8$~cm.
\label{fig4}}
\end{figure}

\begin{figure}           
\plotone{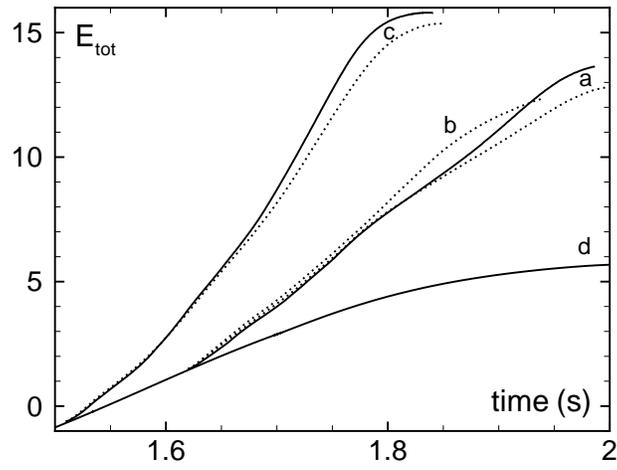}
\caption{
Total energy as function of time for deflagration (d) and 
delayed-detonation cases a, b, c listed in Table~\ref{table1}. 
Dotted lines correspond to low-resolution simulations.
Energy units are $10^{50}$~ergs.
\label{fig5}}
\end{figure}


\end{document}